\documentclass[letterpaper, 10 pt, conference]{ieeeconf}  % Comment this line out
\usepackage{graphics} % for pdf, bitmapped graphics files
\usepackage{epsfig} % for postscript graphics files
\usepackage{mathptmx} % assumes new font selection scheme installed
\usepackage{times} % assumes new font selection scheme installed
\usepackage{amsmath} % assumes amsmath package installed
\usepackage{amssymb}  % assumes amsmath package installed

\title{\LARGE \bf
A Contextual Hierarchical Graph Model for Generating Random Sequences of Objects with Application to Music Playlists
}

\author{Igor de Oliveira Nunes$^{1}$, Gabriel Matos Cardoso Leite$^{1}$, Daniel Ratton Figueiredo$^{1}$\\$^{1}$Programa de Engenharia de Sistemas e Computa\c{c}\~{a}o (PESC/COPPE)\\Universidade Federal do Rio de Janeiro (UFRJ)\\ {\tt\small \{ionunes,gmatos,daniel\}@cos.ufrj.br}
}

\begin{document}

\maketitle
\thispagestyle{empty}
\pagestyle{empty}

\begin{abstract}
Recommending the right content in large scale multimedia streaming services is an important and challenging problem that has received much attention in the past decade. A key ingredient for successful recommendations is an effective similarity metric between two objects, and models that leverage the current context to constrain the recommendations. This work proposes a model for random object generation that introduces two key novel elements: (i) a similarity metric based on the distance between objects in a given object sequence, that is also used to measure similarity between meta-data associated with the objects, such as artists and genres; (ii) a hierarchical graph model with different graphs each associated with a different meta-data. A biased random walk in each graph that are coupled and synchronized dictate the random generation of objects, leveraging the current context to constrain randomness. The proposed model is fully parameterized from sequences of objects, requiring no external parameters or tuning. The model is applied to a large music dataset with over 1 million playlists generating a hierarchy with three layers (genre, artist, track). Results indicate its superiority in generating actual full playlists against two baseline models. 

%With the increasing quantity of readily available multimedia content, such as music, films and books, the development of recommendation systems that facilitate browsing, discovery and playlist construction has become an important and challenging research problem. Many of these systems rely on a notion of distance, or similarity, between such objects. In this work, we propose a new framework to compute distance between objects based on the context of their occurrence in hand-curated playlists. The proposed method is capable of inferring similarity not only between multimedia objects, but also between pairs of other associated entities, such as artists, writers and genres. Using this idea, we design a music playlist generating system based on track, artist and genre similarity networks. The results show....
\end{abstract}

%%%%%%%%%%%%%%%%%%%%%%%%%%%%%%%%%%%%%%%%%%%%%%%%%%%%%%%%%%%%%%%%%%%%%%%%%%%%%%%%
\section{INTRODUCTION}
\label{sec:intro}

The emergence of large streaming services significantly increased the amount of multimedia content consumed by Internet users over the last decade. Moreover, consumption pattern is also changing from a pull-based approach where users actively select the content, to a push-based approach where streaming services continuously send content to users. In both approaches, a significant challenge is selecting the right content for a user, as large scale services often offer millions of objects. Recommendation systems tackle this problem by recommending objects that are more likely to be enjoyed by users. Nevertheless, understanding user tastes and preferences continues to be a difficult task. 

In the context of music, streaming services such as Spotify and LastFM offer users the notion of {\em playlists}, a sequence of songs that have some notion of similarity and cohesion that is likely to be enjoyed by a listener. Such playlists can be crafted manually by artists, experts or listeners, and be fixed, determining a particular song sequence. Users then choose among such playlists, facilitated by a recommendation system as the number of playlists can also be very large. Alternatively, playlists can be generated automatically and dynamically and even be personalized to a given user~\cite{bonnin2015automated}. In such scenario, the user listens to a continuous sequence of songs chosen by a recommendation system.  

Despite the heterogeneity among different techniques for recommending content, a fundamental ingredient is the notion of object similarity~\cite{knees2013survey}. In particular, given two objects (e.g., two songs), provide a number that indicates their relationship or similarity under one or more criteria. One well-explored approach known as \textit{content-based} uses signal processing techniques to analyze the objects to determine their similarity, such as spectrum analysis and time series correlations~\cite{logan2002content}. Another more recent and promising approach known as \textit{context-based} leverages meta-data concerning the object to determine their similarity~\cite{slaney2011web,pichl2017improving}. The potential of \textit{context-based} approaches comes from the continuous generation of meta-data by users. While some meta-data is fixed and inherent to the object (e.g., the artist of a song), other meta-data is constantly being generated by users (e.g., number of users that have listened to two given songs). This information can be leveraged to design more effective measures of similarity that will then drive better recommendation systems. 

While information from various sources can be collected and used to design \textit{context-based} techniques, a powerful source of information are sequences of content consumed by users. Such sequences reveal a user's preference along with a sequential ordering, since content is consumed in such order. In the context of music, playlists embody such sequences and have been leveraged to design a measure of similarity between songs~\cite{pachet2000taxonomy,ragno2005inferring}. This work takes the same approach but generalizes prior work to consider direction and multi-hop influence. In the proposed technique, object similarity decays with the number of objects in between them, in the sense that two objects that follow each other are more similar than otherwise. Moreover, similarity is not symmetric in the sense that an object may closely follow another in a sequence, but the reverse is not necessarily true. 

Another kind of information explored to design \textit{context-based} techniques is meta-data inherent to objects~\cite{knees2013survey}. This can be used to stratify the objects into classes, to better assess their similarity for example. This work leverages this approach but combines it with sequences of objects. In particular, a sequence of objects can be translated into a sequence of attributes associated with the objects, giving rise to multiple sequences for different kinds of meta-data. In the context of music, a playlist can be translated into a sequence of artists (i.e., each song has an artist) or a sequence of genres (i.e., each song has a genre) or any other kind of meta-data, such as language, country or year. The proposed similarity metric can the be applied to such translated sequences to construct a similarity metric for different kinds of meta-data. 

While a effective similarity metric is important, the object recommendation requires a model. A recently proposed and promising approach are graph models, where objects are nodes in the graph and edges encode relationships among them~\cite{ragno2005inferring,bogers2010movie,ueda2018contextual}. Random walks on such graphs are often used to generate or recommend content. The approach here proposed constructs multiple directed weighted graphs, each corresponding to a different meta-data of the objects, where edge weights correspond to similarity between nodes. These graphs form a hierarchy according to the different meta-data. Moreover, biased random walks are placed on each graph and are used to generate content. However, these random walks are coupled and walk synchronously on their respective graphs. Intuitively, context provided by a meta-data graph constrains transitions on other graphs, allowing for more effective object generation. In the context of music, a transition in the genre graph to ``rock" will enforce that the song graph must transition to a song of that genre. 

The main contributions of this work are as follows:
\begin{itemize}
\item 
A novel measure for object similarity that leverages sequences of objects. The measure is not symmetric, capturing the inherent direction of sequences, and captures influences at all gaps, beyond just neighboring.
\item 
The translation of an object sequence to sequences of different kinds of meta-data, for which the same similarity measure can be constructed.
\item
A multiple graph hierarchical model (one for each kind of meta-data) that leverages similarities among the meta-data is proposed to generate sequences. Biased random walks, one on each graph, that walk synchronously and constrain one another are used to generate sequences. 
\item
The proposed model is applied to a large music playlist dataset. The playlists are used to generate three kinds graphs (genres, artists and tracks) that are then used to generate random playlists. The model is fully parametrized from the playlists requiring no external parameters or tuning. Part of the dataset is used for parametrization and the model is evaluated on generating actual (never before seen) playlists, showing superiority against using a simple similarity metric and against using a single graph.
\end{itemize}
While the proposed model has been applied to music objects, it can be used in a myriad of other contexts, as long as objects follow a sequence and have meta-data. For example, in the context of short videos, books, and movies. 

The remainder of this work is organized as follows. Section~\ref{sec:framework} introduces the proposed similarity measure and object generation mode. Section~\ref{sec:data} describes the dataset while Section~\ref{sec:evaluation} presents a characterization of the graphs as well as the results of the model. Last, Sections~\ref{sec:related} and~\ref{sec:conclusion} present a brief discussion of related work and some final remarks. 

%In our framework, we create a generalization of this idea where we consider that not only the immediately consecutive appearance of the song pair contributes to their similarity. Instead, we propose that this contribution is higher for songs that are close in the stream and the similarity decays with the number of songs between the considered pair. The reason for this is that, intuitively, the fact that two tracks are contained in the same authored stream, not necessarily adjacent, should give some information about their relationship. Moreover, another generalization we make is to consider the similarity as a non-symmetric relation. In the context of playlist generation, this allows the system to consider that one track is more likely to appear after another, but not the other way around. A relevant aspect of this graph building strategy is that it provides a general procedure to create similarity networks between any entities contained in hand created lists.
%
%We take advantage of this fact in the architecture of our playlists generation system. We propose the construction of similarity networks between three entities relevant to music playlists: genres, artists and tracks. In order to generate the playlists, the networks interact in a layered manner. A random walk is performed in the genre graph, the result impacts another random walk in the artists graph, that finally influences the steps in the tracks which produces the output.

\section{SIMILARITY MEASURE AND GRAPH HIERARCHY}
\label{sec:framework}

Consider a finite set of distinct objects $\cal{O}$ where each object $o \in \cal{O}$ has $k$ different attributes, namely $a_1(o), \ldots, a_k(o)$. Let ${\cal A}_l$ denote the set of possible values for attribute $a_l$ with $l = 1, \ldots, k$. For example, $\cal{O}$ can be the set of songs in a digital repository (e.g., all songs in Spotify), and the attributes can be the title, singer, and genre of a song. 

Consider a sequence of objects ${\cal S} = o_1, o_2, \ldots$ where each object $o_t$ is an element of $\cal{O}$. In the context of songs, a sequence can be a music playlist constructed by a user. An important assumption in what follows is that such sequence encodes some kind of association or similarity between the objects. In particular, two objects that frequently appear close to each other in the sequence are more strongly related (or similar) than two objects that never appear close to each other. For example, in music playlists, songs that frequently appear together tend to be related or similar in the musical sense. This intuition is leveraged to construct a similarity metric between the objects. Moreover, since objects have multiple attributes, a similarity metric can be constructed for each attribute. 

\subsection{Similarity metric}

Consider two objects $o_i$, $o_j \in \cal{O}$ and a sequence $\cal{S}$. Let $t_k(o)$ denote the time of the $k$-th appearance of object $o$ in $\cal{S}$. In particular, $t_k(o) = \min_{t > t_{k-1}(o)} \{ t | o_t = o \}$ where $t_0(i)=0$.  Consider the set of intervals between appearances of objects $o_i$ and $o_j$ in $\cal{S}$. In particular, for an appearance of $o_i$ at time $t$, let $D(i,j,t)$ denote the set of times that $o_j$ appear after $t$. Thus, when considering the $k$-th appearance of $o_i$, we have $D(i,j,k) = \{ t' | t' > t_k(o_i) , o_{t'} = o_j \}$, with a slight abuse of notation. Note that $D(i,j,k)$ can be empty, which occurs when $o_j$ does not appear after $t_k(o_i)$ in the sequence. 

The similarity between two objects can now be defined in terms of its appearance, as follows:
\begin{equation}
    s(o_i,o_j) = \sum_k \sum_{t \in D(i,j,k)} f(t - t_k(o_i))) , 
    \label{eq:sim}
\end{equation}
where $f$ is a positive but monotonically decreasing function, for example, $f(t) = 1/t$. Note that all appearances of $o_j$ that occur after an appearance of $o_i$ contribute to increase their similarity. However, appearances that are close in the sequence contribute more, since $f$ is assumed to be decreasing. Moreover, note that the metric $s$ is not symmetric and $s(o_i,o_j)$ is likely different from $s(o_j,o_i)$. Also, note that an object can have similarity with itself, but this is not necessarily large. 

The sequence $\cal{S}$ of objects be converted into a sequence of a given attribute of the objects. In particular, when considering attribute $a_l$, we can define ${\cal S}_l = \left( g_l(o_t) \right)$, for $t=1, \ldots$ where $g_l(o)$ returns the value of the $l$-th attribute of object $o$. Thus, ${\cal S}_l$ can then be used to define the similarity between two values of the $l$-th attribute, by simply redefining the notions used in equation~\ref{eq:sim} for objects with their respective notion for attribute values $a_i, a_j \in {\cal A}_l$. Thus, sequence $\cal{S}$ can then be used to determine similarities for all attributes, and thus let $s_l(a_i, a_j)$ denote the similarity between the values $a_i$ and $a_j$ of the $l$-th attribute. 

\subsection{Similarity networks and hierarchy}

The above similarity metric can be used to construct a directed weighted network. In particular, let $G=(V,E)$ denote a directed graph where $V = \cal{O}$ denotes the set of vertices and $(o_i,o_j) \in E$ if $s(o_i, o_j) > 0$ the set of edges. Moreover, the weight of $(o_i,o_j)$ is given by $s(o_i, o_j)$. Let $N^+(o_i)$ denote the set of outgoing neighbors of $o_i$, namely $N^+(o_i) = \{ o_j | s(o_i, o_j) > 0 \}$. Moreover, let $W^+(o_i)$ and $W^-(o_i)$ denote the total outgoing and incoming weight of node $o_i$. Note that the network depends on the sequence $\cal{S}$. 

The same kind of network can be constructed for each attribute. In particular, let $G_l=(V_l,E_l)$ denote a directed graph associated with attribute $l$ where $V_l = {\cal A}_l$ denotes the set of vertices and $(a_i,a_j) \in E$ if $s_l(a_i, a_j) > 0$ the set of edges. Thus, there are $k$ networks, one for each attribute associated to the objects. 

\begin{figure}[h]
    \centering
    \includegraphics[width=0.8\columnwidth]{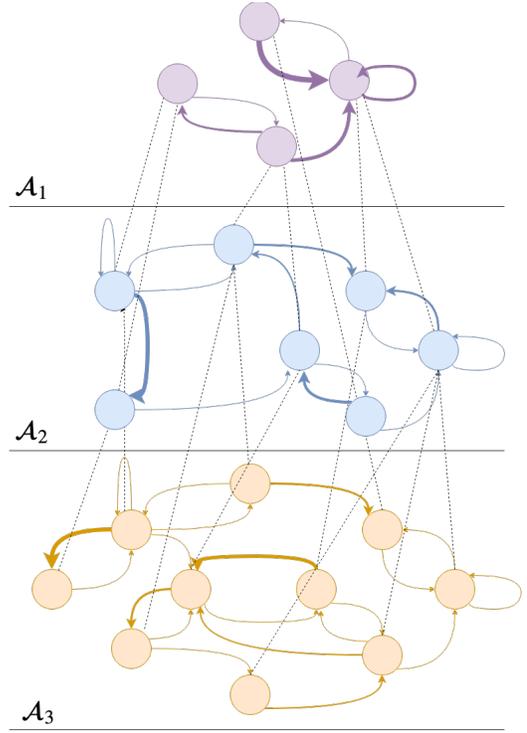}
    \caption{Illustration of a hierarchy with $3$ networks (three layers); dashed lines connect the attributes of given objects; solid line thickness indicate different edge weights (similarities) within a network.}
    \label{fig:illustration}
\end{figure}

These independent networks can be structured into a hierarchy, as follows. Assume that the network size (in number of nodes) increases with the attributes (i.e., assume that the attributes are sorted such that $|{\cal A}_1| < \cdots < |{\cal A}_l| < \cal{O} $). Note that the top of the hierarchy is defined by network $G_1$. Since every object $o$ in the sequence $\cal{S}$ has its attributes defined, $o$ uniquely maps to a node in each network. Thus, the networks are all coupled by objects in the sequence $\cal{S}$. Moreover, note that if an edge $(o_i, o_j)$ exists, then the corresponding edge also exists in all other networks, namely $(g_l(o_i), g_l(o_j) \in E_l$ for all $l=1,\ldots,k$ (by construction of the similarity function). Figure~\ref{fig:illustration} illustrates a scenario where $l=3$. 

\subsection{Sequence generation}

The networks and hierarchy previously defined can be used to generate random biased sequences of objects. The key idea is to leverage the hierarchy to constrain the randomness and thus generate more meaningful sequences. The generation follows biased random walks in each network that are coupled and synchronized, and where the bias is given by edge weights (similarity). In particular, let $X_t = o_i$ denote the location of the random walk in the object network at time $t$. Let $N_t \subset N^+(o_i)$ denote the set of outgoing neighbors of node $o_i$ that are enabled at time $t$ (to be discussed how $N_t$ is determined). Then, the transition probability is given by:
\begin{equation}
    \mathbf{P}[X_{t+1} = o_j | X_t = o_i] = \frac {s(o_i, o_j)}{\sum_{o \in N_t} s(o_i, o)} , \mbox{ if $o_j \in N_t$}
    \label{eq:transition}
\end{equation}
Note that transitions are more likely to objects that are more similar, with a linear dependence between similarity and probability across the set of possible transitions. 

The set $N_t$ is determined according the random walks in the other networks. Consider the random walk in the highest level of the hierarchy, in $G_1$. This random walk is unconstrained and moves freely according to the bias determined by the edge weights. The random walks of all networks are coupled and walk in lockstep. Let $X_1(t), \ldots, X_k(t)$ denote the state of the random walk at time $t$ in each network. In particular, each takes a step after the random walk in the level immediately above has taken a step. Thus, $X_1$ takes a step and moves to $X_1(t+1)$, given this transition, $X_2$ takes a step and moves to $X_2(t+1)$, and so on. The possible transitions in a given layer are constrained by the transition in the layer immediately above. In particular, given $X_1(t+1)$, $X_2$ can only transition to attribute values that have appeared in objects that also have the attribute value given by $X_1(t+1)$. Effectively, this constrains the outgoing neighbors of $X_2(t)$, and thus determines $N_t$. 

In the music example, the top level of the hierarchy can be the genre network, where nodes are music genres. The next level of the network can be the artist network, where nodes are musicians or bands. Finally, the bottom layer of the hierarchy is the track network, where nodes are specific songs. In order to generate a sequence of songs, the random walk in the genre network will make a transition. This will constrain the transitions of the random walk in the artist network, such that only transitions to the genre selected above are enabled. After this transition is made, the random walk in the track network will take a step, but now constrained to the artist selected in the layer above. In particular, only transitions to songs from this artist are enabled. Once this transition occurs, the process repeats and the random walk in the genre network takes a step. Note that the hierarchy constrains the randomness that is further biased by the similarity metric in each network.

\section{DATASET}
\label{sec:data}
In order to illustrate an application of the proposed framework, a dataset of playlists called \textit{Art of the Mix 2011}\cite{mcfee2012_dialect} was chosen. This data is an expansion of the Art of the Mix (AotM)\footnote{http://www.artofthemix.org/} collection of Ellis, et al.\cite{ellis2002} which contains $101343$ unique playlists with varying sizes, each having its own genre such as ROCK, POP, etc. 

\subsection{Augmenting dataset}
Due to the ratio between the number of playlists and the number of distinct tracks in the data ($720100$), it was necessary to augment data without changing much of its characteristics such as the order of tracks appearance. Considering a playlist as a sequence $\cal{S}$, two approaches were employed in the augmentation: randomly remove one element $o \in \cal{S}$ of each playlist and randomly select a pivot to split and recombine the playlist. The first approach leads to a sequence ${\cal{S}}' = {\cal{S}} \setminus o$ while the second leads to a sequence ${\cal{S}}' = \{ o_{p+1}, \dots, o_{n}, o_1, \dots, o_{p-1}, o_p \}$ where $n$ is the size of the sequence $\cal{S}$ and $p$ is the pivot. The first was repeated $9$ times and the second only once, generating a new dataset with $1013430$ playlists, $10$ times bigger than the original.

\subsection{Assigning tracks' genre}
Since the genre provided in the database is associated with a playlist instead of the tracks within it, it is necessary to perform a pre-processing to assign a genre to a track. 

Consider an object $o \in \cal{O}$ with $3$ attributes $a_1(o)$, $a_2(o)$ and $a_3(o)$ defined as genre, artist and track respectively. Let $C_i(o)$ denotes the number of times object $o$ appears in a playlist with genre $i$. The value of $a_3(o)$ is assigned as $a_3(o) = \max\{C_i(o)|\forall i \in {\cal{A}}_3\}$. Among the possible values for genre, there is a value "MIXED GENRE" denoted as $a_{3, MG} \in {\cal{A}}_3$ which is very frequent and adds no information about the real genre of a track and artist. Thus, if $\max\{C_i(o)|\forall i \in {\cal{A}}_3 \setminus a_{3, MG}\} \neq \emptyset$, $a_3(o)$ is assigned as $a_3(o) = \max\{C_i(o)|\forall i \in {\cal{A}}_3 \setminus a_{3, MG}\}$. After this, the size of each set was $|{\cal{A}}_1| = 43$, $|{\cal{A}}_2| = 174566$ and $|{\cal{A}}_3| = 720100$.

\section{EVALUATION}
\label{sec:evaluation}

\subsection{Network characterization}

\begin{figure}[h]
    \centering
    \includegraphics[width=\columnwidth]{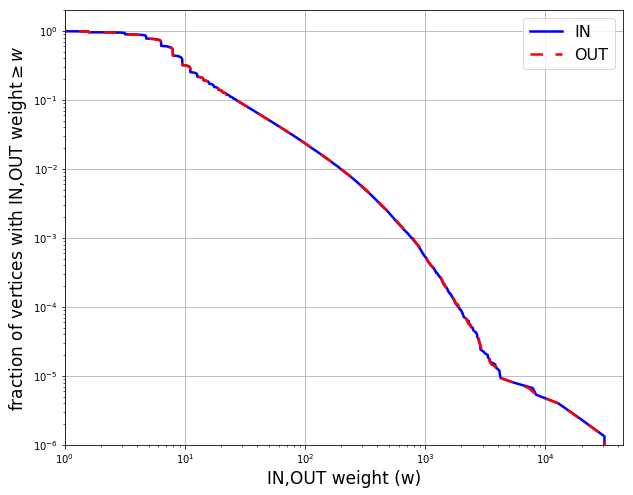}
    \caption{Empirical complementary cumulative distribution (CCDF) of the sum of incoming and outgoing edge weights for vertices in the track graph.}
    \label{fig:in_out_weight_ccdf}
\end{figure}

The networks generated according to the proposed models present specific topological characteristics regarding weights distribution and the influence that different decaying functions have on it. Figure \ref{fig:in_out_weight_ccdf} shows the distribution of the sum of each vertex's incoming and outgoing edge weights. Although there is no significant difference between both distributions, it is possible to notice that a very small number of vertices have very large associated weights. This means that there are some popular songs that appear very often in playlists.

\begin{figure}[h]
    \centering
    \includegraphics[width=\columnwidth]{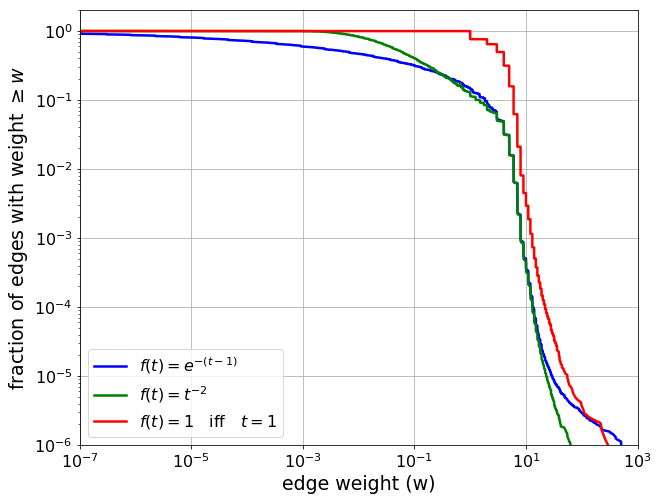}
    \caption{Empirical complementary cumulative distribution (CCDF) of edge weights in the track graph when using different definitions for $f(t)$.}
    \label{fig:edge_weight_ccdf}
\end{figure}

In Figure \ref{fig:edge_weight_ccdf} we present the influence of the decaying function $f(t)$ of Equation \ref{eq:sim} in the edge weights distribution. This model parameter allows to choose how much importance is given to distant objects in the sequences. 

Another important structural aspect -- known as Giant Connected Component -- is present in our networks as it is in many other real world complex networks. In the tracks graph, $89.4\%$ of the vertices are in the largest connected component; in the artists graph, $95.6\%$ and the genre graph all vertices are in the same connected component.

\subsection{Playlist evaluation}

The playlists dataset was split in train and test in three different proportions: $[50\%,50\%]$, $[70\%,30\%]$ and $[90\%,10\%]$. Also, two models were built using the proposed framework: one with 3 hierarchy levels (genres, artists, tracks) named \textit{hierarchical} and the other with a single level (tracks) named \textit{multi hop}, both using $f(t) = e^{-(t-1)}$. They were also compared to the model proposed in \cite{ragno2005inferring}, named \textit{single hop}, which uses an undirected, single layered track network where
\begin{equation}
    {f(t) = \begin{cases}1, \text{ if }t = 1\\ 0, \text{otherwise}\end{cases}}
    \label{eq:f1}
\end{equation}
In order to compare the three models, we used a language modelling inspired approach, proposed in \cite{mcfee2011natural}, which evaluates how likely each model is to produce naturally occurring playlists according to:

\begin{equation}
    {\cal{L}}({\cal{P}} | M) = \frac{1}{|{\cal{P}}|} \sum_{{\cal{S}} \in {\cal{P}}} \log{\mathbf{P}_{M}[{\cal{S}}]}
    \label{eq:loglikelihood}
\end{equation}

 where $M$ is the model, ${\cal{P}}$ is a sample of playlists from the test set and $\mathbf{P}_M[\cal{S}]$ is the likelihood of playlist $\cal{S} \in \cal{P}$ being generated by the model $M$. Thus, we can say that $M_{1}$ is a better model of the data than $M_{2}$ if ${\cal{L}}({\cal{P}} | M_{1}) > {\cal{L}}({\cal{P}} | M_{2})$.
 
 The likelihood $\mathbf{P}_M[\cal{S}]$ is defined as 
 
 \begin{equation}
    \mathbf{P}_M[{\cal{S}}] = \prod_{(o_i, o_j) \in {\cal{S}}} \mathbf{P}[X_{t+1} = o_j | X_t = o_i]
    \label{eq:likelihood}
\end{equation}

\subsubsection{Smoothing}
\begin{figure}[h]
    \centering
    \includegraphics[width=\columnwidth]{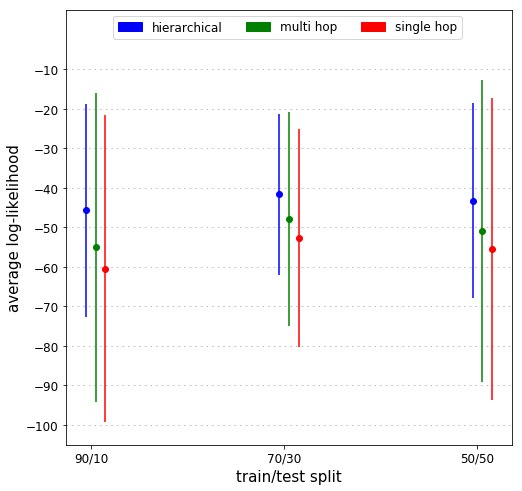}
    \caption{Average log-likelihood of test playlists obtained by each model considering different train/test splits.}
    \label{fig:ALL}
\end{figure}

When evaluating the probability of a sequence of objects $o_i$, $o_j$, it may occur that $s(o_i, o_j) = 0$ in which case $\mathbf{P}[X_{t+1} = o_j | X_t = o_i] = 0$. In this case, $\mathbf{P}_M[\cal{S}]$ does not yield any meaningful value, as the likelihood of this sequence will be $0$ since one transition in the sequence is not possible. In order to allow the computation of the likelihood for sequences with a non-existing transition, the sequence probability was modified and inspired by Laplace Smoothing used in Natural Language Processing. This new sequence probability is defined as follows:
\begin{align*}
    \mathbf{P}&[X_{t+1} = o_j | X_t = o_i] = \frac {s(o_i, o_j) + \frac{1}{|{\cal{A}}_k|}}{\sum_{o \in N_t} s(o_i, o) + \frac{N_t}{|{\cal{A}}_k|}} , \\
    &\mbox{ for $o_i, o_j \in {\cal{A}}_k$}
    \label{eq:smoothing}
\end{align*}
Note that when $s(o_i, o_j) = 0$, the numerator becomes the constant $1/|{\cal A}_{k}|$. The obtained results for each model considering the different splits of the train and test are shown in Figure \ref{fig:ALL}.

The result shows that the proposed model (multi hop) outperforms the single hop by at least five orders of magnitude. This indicates the effectiveness of the proposed generalization to consider not only immediate adjacencies in the track sequence, but also generating transitions between two songs that are not neighbors in the playlist. Moreover, when constrained by the hierarchical structure, the model achieves even better results, confirming the hypothesis that the objects' inherent hierarchy is important for sequence generation.

\section{RELATED WORK}
\label{sec:related}

Recommendation systems for multimedia objects is an important research area that has been widely explored over the last decades~\cite{bobadilla2013recommender}. Within this area, music recommendation has also received much attention recently~\cite{bonnin2015automated,knees2013survey}, in part due to the emergence of large scale streaming services, such as Spotify and lastFM. Despite significant progress, music recommendation continues to be a hard problem, and was the theme of a challenge at the ACM Conference on Recommender Systems (RecSys) in 2018~\cite{chen2018recsys}. 

While there is a large number of approaches and algorithms for multimedia recommendation, they often fall into one of two broad categories: {\textit content-based}~\cite{logan2002content,su2010music} and {\textit context-based}~\cite{slaney2011web,pichl2017improving,knees2016collaborative}. Within the {\textit context-based} approach there are various graph-based techniques where a graph of objects is constructed and then leveraged for recommendations, often using random walks~\cite{bogers2010movie,ragno2005inferring,ueda2018contextual}. 

From a theoretical perspective, Gopal et al.~\cite{gopal2012bayesian} discuss many challenges in the  hierarchical classification problem and propose a Bayesian hierarchical model using multivariate logistic regression. Based on this idea, Ben-Elazar et al.~\cite{ben2017groove} proposed an algorithm that is used in Microsoft's Groove music service. The approach leverages a variational Bayes technique for learning the parameters of a hierarchical model that integrates genre, sub-genre, artist and global information while also incorporating personalization for user-specific preferences.

The approach here proposed is closely related the methodology proposed by Ragno et al.~\cite{ragno2005inferring} where an undirected weighted graph is constructed from playlists and a biased random walk is used to generate recommendations. The proposed similarity metric simply counts the number of times two songs have appeared next to each other in the playlists, and this is used as edge weights. This current work generalizes this methodology by introducing multiple graphs that induce a hierarchy and a similarity metric that is not symmetric (directed graph) and leverages multiple hops between two objects in a sequence of objects. This current work is also closely related to the recent work of Ueda et al.~\cite{ueda2018contextual} where a single directed graph with nodes that represent different objects such as tracks, genres and artists is constructed. Their methodology uses a single biased random walk to generate recommendations, but this bias is an external parameter. Moreover, object similarity is not considered in the construction of the graph and recommendation requires extensive computation (in order to compute the next most likely track). The approach here proposed more naturally encodes the relationships between the different kinds of meta-data (in a hierarchy) that is then used to constrain random walk transitions. Moreover, this approach is quite general and can be trivially applied to other kinds of objects, beyond music.

\section{CONCLUSION}
\label{sec:conclusion}

Making good recommendations for multimedia objects is an important and challenging task that continues to draw attention from academia and industry despite over a decade of progress. The many approaches that have emerged in the literature indicate that successful recommendations require using effective similarity metric that assess the similarity between two objects as well as models that leverage current context to constrain the recommendations. This work proposed a novel approach to both ingredients. 

The proposed similarity measure is constructed from just a sequence of objects and assigns similarity inversely proportional to the distance between objects in the sequence. More appearances leads to a larger similarity, as well as closer appearances in the sequence. This notion can be applied to any meta-data associated with the objects (giving rise to a sequence of meta-data values) effectively constructing a suite of similarity measures. 

The proposed model consists of a hierarchy of graphs each corresponding to a meta-data of the objects. The weights of these directed graphs correspond to similarity between the meta-data (and is not symmetric). In order to generate a recommendation, each graph has a random walk that move coupled and synchronously. Thus, a transition in a given layer of the hierarchy constrains the possible transitions in the layer immediately below, which in turn will constrain the transitions in the layer below it. 

The proposed model was applied to a music dataset containing 1 million playlists where a hierarchy with three layers was constructed: genre, artist, and track. Results indicate that the model can generate actual (never before seen) playlists with an accuracy that is at least 5 orders of magnitude higher then two alternative approaches. 

Last, although this model has been applied to the context of music, it could also be applied to any other context as long as objects in a sequence exert some notion of cohesion with respect to their distances in the sequence and also have meta-data associated with them. Examples, are short videos or books, but further analysis is needed to assess this generality.

\addtolength{\textheight}{-12cm}   % This command serves to balance the column lengths

%%%%%%%%%%%%%%%%%%%%%%%%%%%%%%%%%%%%%%%%%%%%%%%%%%%%%%%%%%%%%%%%%%%%%%%%%%%%%%%%

%%%%%%%%%%%%%%%%%%%%%%%%%%%%%%%%%%%%%%%%%%%%%%%%%%%%%%%%%%%%%%%%%%%%%%%%%%%%%%%%
%\section*{APPENDIX}

%\section*{ACKNOWLEDGMENT}
%
%The preferred spelling of the word ÒacknowledgmentÓ in America is without an ÒeÓ after the ÒgÓ. Avoid the stilted expression, ÒOne of us (R. B. G.) thanks . . .Ó  Instead, try ÒR. B. G. thanksÓ. Put sponsor acknowledgments in the unnumbered footnote on the first page.

%%%%%%%%%%%%%%%%%%%%%%%%%%%%%%%%%%%%%%%%%%%%%%%%%%%%%%%%%%%%%%%%%%%%%%%%%%%%%%%%

%References are important to the reader; therefore, each citation must be complete and correct. If at all possible, references should be commonly available publications.

\bibliographystyle{IEEEtran}
\bibliography{references}

\end{document}